# Quantum Process in Living Cells

ROBERT W. FINKEL
Physics Department
St. John's University
Jamaica New York 11439
USA
finkelr@stjohns.edu

*Abstract:* - Quantum processes have been confirmed for various biological phenomena. Here we model a quantum process in cells based on coherent waves of established ultrafast energy transfers in water. We compute wave speed, ~156 km/s, and wavelength, ~9.3 nm, and determine that the waves retain local coherence. The model is compared with observations and diverse numerical applications lend support to the hypothesis that rapid energy transfers in water are characteristic of living cells. Close agreements are found for the dipole moment of water dimers, microwave radiation on yeast, and the Kleiber law of metabolic rates. We find a sphere with diameter ~20 nm is a lower bound for life in this theory. The quantum properties of the model suggest that cellular chemistry favors reactions that support perpetuation of the energy waves.

## 1. Introduction

Coordinated chemical reactions are characteristic of living cells. Since quantum physics determines the cooperatively of various many-body systems like superconductors, superfluids, and solid-state materials, one may premise that quantum physics also plays a cooperative role in biology. Schrödinger posited that quantum effects might distinguish living from non-living matter revealing "new" physical behavior [1] and Fröhlich advocated that all living cells maintain coherent vibrations generated by metabolism [2]. Investigations have since confirmed that quantum processes exist in several biological phenomena [3]. We embrace the concept of a cooperative quantum process in cells and explore a model mechanism that reproduces several empirical results.

The cooperative process proposed here originates from waves of energy propagating through the water matrix of cells. For our present purposes it is not necessary to identify a metabolic origin for this energy. Various features of cellular organization have been attributed to the collective dynamics of water [4]. Water has long been known to transfer energy at an ultrafast rate following Woutersen and Bakker [5]. The agents of these energy transmissions are water dimer oscillations that spread by successive excitations of dipole–dipole interactions (a Förster transfer mechanism) [5,6]. According to the matter-wave duality of quantum physics, the resulting energy waves have corresponding particle quanta here termed *chaions*.

### 1.1 Background

Quanta of the ultrafast energy transfers do not convey mass so chaion particles are analysed using a statistical treatment similar to that for a massless photon gas. This treatment is detailed in reference [7] and here we summarize results needed for our applications.

Integrating the occupation number over phase space gives the average number of energy-carrying particles, $\langle n \rangle$,

$$\langle n \rangle = 1.202 \pi V \left( \frac{k_B T}{h v} \right)^3, \qquad (1)$$

Where $v$ is the particle velocity, $h$ is Planck's constant, $k_B$ is Boltzmann's constant, $T$ is temperature, and $V$ is the volume of water. Similarly, integrating chaion energies occupied over all of phase space gives the average energy vested in the quanta with the result that the average energy per chaion particle is $2.701 k_B T$. Hereafter we abbreviate the average energy per particle as $2.7 k_B T$ while using the more precise value in computations. Frequency $f$ is given as usual by $f = 2.7 k_B T / h$ and the energy density $\rho$ is uniform throughout the volume.

## 2. Wave properties

We now determine chaion velocity and wavelength directly from the model. Although relativity requires that massless particles in vacuo must propagate at the speed of light, $c$, water transmits energy at a nonrelativistic speed, $v$. This is easily reconciled if excited water dimers transmit energy successively with probability $p$ such that the effective speed is a fraction of light speed $p = v/c$ (we neglect the transit time between dimers).

Consider an assembly of $N$ water molecules containing one excited dimer able to emit its energy of $2.7 k_B T$ with probability $v/c$. Let $V_w$ represent the volume of a spherical water molecule with effective diameter 3.2 Å [8]. Equation (1) becomes

$$\frac{v}{c} = \pi \zeta(3) N V_w \left( \frac{k_B T}{h \nu} \right)^3$$

Alternatively, the probability of a single emission is also expressed by the canonical distribution for the dimer energy attributed to one molecule.

$$\frac{v}{c} = \frac{\exp(-2.7)}{(N-1) + \exp(-2.7)}$$

The latter two equations are solved for $N$ and $v$ at standard temperature giving $N = 129$ and $v = 156$ km/s. (This speed is virtually that of an electron excited with chaion energy $2.7 k_B T$.) The wave relation $\lambda f = v$ now determines the characteristic wavelength as $\lambda = 9.3$ nm.

### 2.1 Evade Decoherence

Although a warm wet cell seems an unlikely place for coherent quantum processes, the waves considered here are sufficiently fast and short enough to avoid decoherence. The characteristic times considered in the chaion theory are about $f^{-1} = 59$ fs while biochemical reaction times are larger by orders of magnitude. The measured equilibration time for pure water is 0.55 ps [9] exceeding chaion transit time almost tenfold. Energy waves therefore "see" their environment as being relatively time-independent.

The chaion model presupposes that quantum coherence extends over wavelength $\lambda$ in the warm wet cell. Coherent quantum behavior requires that the time for decoherence $\tau$ exceeds the characteristic chaion transfer time. An approximation for massless particles with relatively long wavelengths and high temperature is given by Schlosshauer [10],

$$\tau = \frac{1}{\Lambda (\Delta x)^2} \quad (2)$$

where $\tau$ is the characteristic time to decohere in a distance $\Delta x$ by a factor of e and $\Lambda$ is a scattering constant,

$$\Lambda = 8! \frac{8 a^6}{9 \pi} 1.002 \left( \frac{k_B T}{\hbar v} \right)^9 v,$$

parameterized by wave speed, temperature, and water molecule size $a = 3.1 \times 10^{-10}$ m. Substituting the chaion quantities $v$ and $T = 310$ K with $\Delta x = \lambda$ in Eq.(2) gives $\tau = 1.3 \times 10^{-12}$ s, two orders of magnitude longer than the chaion transfer time.

The plausibility of quantum coherent states existing in biological systems has been addressed using fundamental considerations [11]. In contrast, our approach here was to accept the existence of coherence at the outset and then explore the consequences of the assumption. Now we see after the fact that our proposed mechanism does avoid decoherence.

## 3. Applications

The model is now tested against experimental data. We find accurate numerical agreements for the dipole moment of water dimers, microwave radiation on yeast, and the Kleiber law of metabolic rates. The computations use only the model quantities and standard physical values with a minimum of mathematics.

## 3.1 Dipole Moment

Energy transmission in water is passed along water dimers. Consider an isolated dimer that can emit only radiant energy. The average power expended by this excited dimer is the product of energy $2.7k_BT$ and the emission probability $p$ divided by the time for emission $f^{-1}$. Set this equal to the Lamor expression for the total average power emitted by a single excited dimer to obtain,

$$2.7k_BT\, p\, f = \frac{q^2\omega^4}{12\pi K\varepsilon_0 c^3}$$

where $K\varepsilon_0$ is the dielectric constant for water at 25°C (K=78.54), $\omega$ is the angular frequency, and $q$ is the dipole moment. All the quantities except for $q$ are known from the chaion model and water properties. Solving gives $q$=2.66D in excellent agreement with the experimental value 2.64D [12].

## 3.2 Microwave Experiments

In a series of experiments, Grundler and co-workers [13-15] subjected yeast cells (*Saccharomyces cerevisiae*) to frequencies in the gigahertz region. They found that cell growth exhibited resonant changes of 10 to 20 percent at frequencies neighboring 41.7 GHz and determined that the effect is not thermal.

We expect that the radiation excites mechanical energy waves that can interfere with chaion energy waves. Constructive interference is then expressed as

$$\lambda_s = n\lambda$$

where $\lambda_s$ represents the wavelength of the sound wave, $n$ is a small integer, and $\lambda$ is chaion wavelength. The incident microwaves of frequency $f_\gamma$ excite sound waves with matching frequency. In terms of $f_\gamma$ and the speed of sound, $v_s$, applying the wave relation $\lambda f = v$ to the latter expression gives a relation for the critical frequency,

$$f_\gamma = \frac{1}{n}\frac{v_s}{v}f$$

where the un-subscripted quantities apply to chaion properties evaluated at the experimental temperature 31°C and the standard speed of sound in soft tissue is 1540 m/s [16]. The integer must be chosen to give a match within a broad absorption peak of water between 40 GHz and 100 GHz [17] necessitating $n=4$. This yields a frequency of 41.6 GHz in agreement with the experimental mean value of 41.7 GHz.

The microwave experiments also showed satellite periodicities in growth recurring in approximately 10 MHz intervals in the critical region from 41.5 to 41.9 GHz. These periodicities can be estimated using the current model. Compton scattering from electrons in the cell perturb part of the incident wave and the greatest displacement from critical frequency occurs when a microwave photon is backscattered. The backscattered frequency of the sound wave is reduced by 5.5 MHz from peak to trough or 11 MHz per cycle in reasonable accord with experiment.

In principle, the calculation can be repeated for successive updated values of frequency $f_\gamma$ producing a series of satellite resonances with diminishing energy. Again, this is consistent with the experimental results.

## 3.3 Kleiber Law

The Kleiber law [18,19] is an empirical rule expressing that an organism's metabolic rate is proportional to the ¾ power of its mass. Here the expression is derived—including the coefficient—from general thermal considerations and the chaion model. Consider a cell to be comprised of thin nested shells. The outer surface is bounded by a thin shell of volume $dV = Avdt$ where $A$ is the cell surface area and $dt$ is an increment of time. This mathematical construct treats the cell as a uniform cytosol with no particular recognition of cell membranes or organelles. We focus on the outer shell where energy is lost to the outside.

Energy entering the outer shell in time $dt$ is the product of energy density $\rho$ and $dV$ with a geometric factor of 1/4 giving $\frac{1}{4}\rho dV$. The one-quarter factor is familiar in derivations of Stefan-Boltzmann radiation from thermal energy density. This detail is elaborated in Ref.[7], but elementary arguments cite a factor of 1/2 arising from half the particles being directed toward the surface and another 1/2 from the average component of particle velocity normal to the surface being $v/2$.

Let $W$ represent the power expenditure generated by the whole enclosed volume so that the fraction of power generated in the outer shell is $dW/W$ and the energy generated within the shell is $\rho V\, dW/W$. The total energy in the shell, $\rho dV$, is the sum of energy generated within the shell and entering the shell

$$\rho dV = \rho V \frac{dW}{W} + \frac{1}{4}\rho dV$$

A solution to this differential equation is $W = aV^{\frac{3}{4}}$ where $a$ is a constant of integration. Converting volume to equivalent mass $M$ introduces a new coefficient $b$ and gives the standard Kleiber form $W = bM^{\frac{3}{4}}$.

Although a single measurement can determine coefficient $b$, it is supportive to calculate it from first principles based on the chaion theory. Choose the special case of a single chaion occupied by volume $V$. The volume $V$ containing an average of one chaion is readily found from Eq.(1) and $M$ is determined by choosing specific cell density to be 1.1 and evaluating all quantities at 37°C. The outwardly directed linear energy transfer, $\mu$, (energy per length) is $2.7 k_B T$ per wavelength so the power emitted is $W = \mu v = 2.7 k_B T v$. Substitute this into $W = bM^{\frac{3}{4}}$ to find $b = 3.4$ in precise agreement with the empirical value for the Kleiber law expressed in Watts, $W = 3.4 M^{3/4}$.

The derivation is readily extended to multicellular organisms when the idealizations of uniform temperature, density, and homogeneity hold to some approximation. Clearly, anomalies will occur when tissues or organs vary significantly in water content or energy and mass densities.

Interest in the Kleiber law was renewed by derivations based on nutrient distribution systems [20,21] that established the ¾ factor. Our treatment complements these and offers an idealized thermal view with the benefit of producing the observed coefficient. The Kleiber law has long been controversial especially because it was alleged to be universally applicable–which it is not. Naïve scaling suggests a ⅔ power law and the actual exponent often diverges from the expected ¾. Nevertheless, the ¾ exponent is a consensus average [22]. Our model produces ¾ but is starkly idealized even for a single cell. We have seen that many further idealizations are needed to apply it to approximate a multicellular organism.

## 4. Discussion and Conclusion

This note posits that ultrafast energy transfers in water are a basis for cooperative properties in living cells. Direct applications of the model accurately determined disparate findings for microwaves on yeast, the Kleiber law, and the dipole moment of water dimers. It is noteworthy that no free parameters were introduced in these successful applications.

The model also predicts wave speed and wavelength as well as the energy of the associated particles. These are subject to experimental confirmations.

The theory implies that chaion waves tend to be self-perpetuating. Harmonic perturbations induce transitions in the neighborhood of a resonant frequency. As a familiar example, photons matching the resonant frequency of target molecules cause cascades of coherent light by stimulated emission. Similarly, chaions match the resonant frequency of excited water dimers and induce further chaion emissions. Such emissions are more probable than competing reactions and the quantum system favors this state over a multitude of possibilities. The system can then select the best choices for its perpetuation perhaps including enzyme-substrate matches or metabolic reactions that can support chaion production.

In this view the living cell favors chemistry that sustains chaion production. Proto-life forms can be imagined to select primitive enzymes and even coding systems without having to assemble the separate parts serially. Patel [23] has argued that the universal coding system of 4 nucleotide bases and 20 amino acids results from an optimum quantum database search according to the Grover algorithm [24]. This seems somewhat plausible in a quantum environment on the order of ~9 nm. A sphere enclosing a single chaion is readily found from Eq.(1) to have a diameter

~20 nm giving a lower bound for life in this theory. Nanobes this small are observed [25], but are not universally accepted to be living.

Although the hypothetical consequences are intriguing, the model must be supported by quantitative results. These are consistent with the premise that cellular life is characterized by waves of rapid energy transfers in water and this process can assist in the chemical organization and coherence of the cell. We conclude that further study of the model is warranted.